\documentstyle[multicol,aps,prl]{revtex}
\begin{document}
\tightenlines
\title{\bf{ Nonequilibrium Growth problems}}
\author{Sutapa Mukherji\cite{eml1}} 
\address{ Department of Physics and Meteorology, Indian Institute of
  Technology, Kharagpur 721 302}
\author{ Somendra M. Bhattacharjee\cite{eml2}} 
\address{Institute of Physics, Bhubaneswar 751 005}
\date{\today}
\maketitle
\widetext
\begin{abstract}
We discuss the features of nonequilibrium growth problems, their scaling 
description and their differences from equilibrium problems.  The
emphasis is on the Kardar-Parisi-Zhang equation and the renormalization 
group point of view.  Some of the recent developments along these
lines are mentioned.
\end{abstract}

\begin{multicols}{2}

\section{Introduction}
  
How to characterize the degree of roughness of a surface as it grows
and how the roughness varies in time have evolved into an important
topic due to diverse interest in physics, biology, chemistry and in
technological applications.  One crucial aspect of these
nonequilibrium growth processes is the scale invariance of surface
fluctuations similar to the scale invariance observed in equilibrium
critical point phenomena.  Although different kinds of growths may be
governed by distinct natural processes, they share a common feature
that the surface, crudely speaking, looks similar under any
magnification and at various times.  This nonequilibrium
generalization of scaling involving space and time (called ``dynamic
scaling'') makes this subject of growth problems important in
statistical mechanics.
  
Growth problems are both of near-equilibrium and nonequilibrium
varieties, and, therefore, they provide us with a fertile ground to
study the differences and the extra features that might emerge in a
nonequilibrium situation\cite{tim,barbasi}.  Take for example the case
of crystal growth.  In equilibrium, entropic contributions generally
lead to a rough or fluctuating surface, an effect called thermal
roughening, but, for crystals, because of the lattice periodicity, a
roughening transition from a smooth to rough surface, occurs at some
temperature.  The nature of the growth of a crystal close to
equilibrium expectedly depends on whether the surface is smooth or
rough.  One can also think of a crystal growth process which is far
away from equilibrium by subjecting it to an external drive, for
instance by random deposition of particles on the surface.  The
roughening that occurs in the nonequilibrium case is called kinetic
roughening.  Is the nature of the surface any different in kinetic
roughening?  Crystals are definitely not the only example of growth
processes; some other examples of such nonequilibrium growths would be
the growth of bacterial colonies in a petri dish, sedimentation of
colloids in a drop, the formation of clouds in the upper atmosphere,
and so on. Note the large variation of length scales of these
problems.  In many such examples it is difficult if not impossible to
think of an equilibrium counterpart.

Scale invariance in interface fluctuations implies that fluctuations
look statistically the same when viewed at different length scales.  A
quantitative measure of the height fluctuation (height measured from
an arbitrary base) is provided by the correlation function 
\begin{equation}
C({\bf x},t)=\langle[h({\bf x}+{\bf x}_0,t+t_0)-h({\bf x}_0,t_0)]^2\rangle,
\label{eq:4}
\end{equation} 
where $x$ and $t$ denote the $d$ dimensional coordinate on the
substrate and time respectively.  The averaging in Eq. (\ref{eq:4}) is
over all ${\bf x}_0$, and, by definition, $C({\bf x},t)$ is
independent of the choice of the arbitrary base.  In simple language,
scale invariance then means that when the system is, say, amplified by
a scaling $x\rightarrow bx$ and $t\rightarrow b^z t$, the height
fluctuations reveal the same features as the original, upto an overall
scale factor.  Quantitatively, there exists a generalized scaling
\begin{eqnarray}
C(x,t)=b^{-2\chi}C(bx,b^zt),\label{scaling}
\end{eqnarray}
where $b$ is a scale factor, and $\chi$ and $z$ are known as the
roughening and dynamic exponents which are also universal.  As a
direct consequence of (\ref{scaling}), a scaling form for $C(x,t)$ can
be obtained by choosing $b=1/x$
\begin{eqnarray}
C(x,t)=x^{2\chi}\hat C(t/x^z),
\label{eq:3}
\end{eqnarray}
a form that also explains the origin of the name ``dynamic exponent''
for $z$. The power law behaviour (as opposed to say exponential decay)
of the correlation function implies absence of any scale, neither in
space nor in time. All the underlying length scales required to define
the problem dropped out of the leading behaviour in Eq. \ref{eq:3}
Such a scale invariance is one of the most important features of
equilibrium phase transitions and is observed when a parameter, say
the temperature, approaches its critical value. However, here there is
no special tuning parameter; the scale invariance appears from the
interplay of competing processes which in the simplest case can be the
surface tension and noise present due to inherent randomness in the
growth.  There can be, of course, more complex events like a phase
transition between surfaces with different roughness but scale
invariance (not only of the correlation function but of any physical
quantity) is generically preserved in all these surfaces.

It is worth emphasizing the enormous simplification that occurs in the
scaling description.  It is only a very few quantities that define the
asymptotic behaviour of the system.  Consequently, the idea of
studying the universal aspects of growth processes is to classify and
characterize the various universality classes as determined by the
exponents, e.g., $\chi$ and $z$, the scaling function and if necessary
certain other important universal quantities.

At this point, it might be helpful to compare the equilibrium and
nonequilibrium cases again.  In equilibrium, thanks to thermal energy
(or ``random kicks'' from a heat reservoir), all configurations of a
system are accessible and do occur, but no net flow of probability
between any two states is expected (called ``detailed balance'').
Consequently, the knowledge of the states (and the energies) of a
system allows one to obtain the thermodynamic free energy by summing
over the Boltzmann factors $\exp(-E/k_BT)$, where $E$ is the energy of
the state, $T$ the temperature and $k_B$ is the Boltzmann constant.
In a nonequilibrium situation, either or both of the above two
conditions may be violated, and the framework of predicting the
properties of a system from free energy is not necessarily available.
A dynamic formulation is needed.  By assigning a time dependent
probability for the system to be in a configuration at a particular
time, one may study the time evolution of the probability. The
equilibrium problem can be viewed from a dynamical point also.  This
description must give back the Boltzmann distribution in the infinite
time steady state limit. This is the Fokker Planck approach.  The
probabilistic description comes from the ensemble picture where
identical copies of the same system exchange energy with the bath
independently.  An alternative approach which finds easy
generalization to the nonequilibrium cases is the Langevin approach
where one describes the time evolution of the degrees of freedom, in
our example $h({\bf x},t)$, taking care of the random exchange of
energy by a noise.  The dynamics we would consider is dissipative so
that the system in absence of any noise would tend to a steady state.
However for it to reach the equilibrium Boltzmann distribution in the
presence of noise, it is clear that the noise must satisfy certain
conditions (Einstein relation) connecting it to the system parameters.
The nonequilibrium case does not have any thermodynamic free energy as
a guiding light and therefore, there is no requirement to reach the
Boltzmann distribution.  In the Langevin approach, the noise term can
be completely independent.  In the equilibrium case the Langevin
equation will be determined by the Hamiltonian or the free energy of
the system, but for nonequilibrium cases there might be terms which
cannot be obtained from Hamiltonians.  Since for $t\rightarrow\infty$,
the probability distribution for equilibrium cases attains the
Boltzmann distribution, the roughness exponent $\chi$ is determined
even in dynamics by the stationary state while the details of the
dynamics is encoded in the dynamic exponent $z$.  In other words the
two exponents $\chi$ and $z$ are independent quantities.  In the
nonequilibrium case, there is no compulsion to reach any predetermined
stationary state and therefore the surface roughness is related to the
growth, i.e., $\chi$ and $z$ need not be independent.  We see below
that there is in fact a specific relation connecting these two
exponents.

The existence of scale invariance and universal exponents implies that
as far as the exponents are concerned the theory should be insensitive
to the microscopic details or in other words one may integrate out all
the small length scale features. The universal exponents come out as
an output of this process of coarse graining of say the Langevin
equation, followed by a length rescaling that brings the system back
to its original form. The new system will however have different
values of the parameters and one can study the flow of these
parameters in the long length and time scale limit.  This is the basic
idea behind the renormalization group (RG).  In this approach the
importance of an interaction or a term is judged not by its numerical
value but rather by its relevance.  One may start with any physically
possible process in dynamics and see how it appears as the length
scale or resolution changes. For large length scales, one is left only
with the relevant terms that grow with length, and marginal terms that
do not change; the irrelevant terms or interactions that decay with
length scale automatically drop out from the theory. The exponents are
determined at the fixed points of the flows in the parameter space.
These fixed points, which remain invariant under renormalization,
characterize the macroscopic or asymptotic behavior of the system.
Clearly from this view point of renormalization group, one can explain
why the microscopic details can be ignored and how the idea of
``universality'' emerges.  All systems whose dynamical behaviour would
flow to the same fixed point under RG transformation will have
identical scaling behaviour.  The various universality classes can
then be associated with the various fixed points of RG
transformations, and phase transitions or criticality with unstable
fixed points or special flows in the parameter space.  An RG approach
therefore seems rather natural and well suited for studying any scale
invariant phenomena in general, growth problems in particular.  Quite
expectedly, the modern approach to growth problems is based on these
views of RG.

For a quantitative discussion, we consider two simple equations that,
from the historical point of view, played a crucial role in the
development of the subject in the last two decades.

A simple Langevin equation describing the dynamics of a surface is the
Edwards-Wilkinson (EW) equation\cite{ew82}
\begin{eqnarray}
\frac{\partial h}{\partial t}=\nu\nabla^2 h+\eta(x,t),\label{ew}
\end{eqnarray}
where $x$ represents in general the coordinate on the $d$ dimensional
substrate. $\nu$ is the coefficient of the diffusion term trying to
smoothen the surface and $\eta$ is the Langevin noise which tries to
roughen the surface\cite{ew82}. One may add a constant current $c$ to
the right hand side, but by going over to a moving frame of reference
($h \rightarrow h+ct$) one recovers Eq. \ref{ew}.  The noise here is
chosen to have zero mean and short range correlation as $\langle
\eta({\bf x},t)\eta({\bf x}',t')\rangle=2D \delta({\bf x}-{\bf
  x}')\delta(t-t')$.  One of the important assumptions in this
equation is that the surface is single-valued and there are no
overhangs.  One can solve (\ref{ew}) exactly just by taking the
Fourier transform and obtain the exponents $\chi=(2-d)/2$ and $z=2$.
That the dynamic exponent $z$ is 2 follows from the simple fact that
the equation involves the first derivative in time but a second
derivative in space.  The surface is logarithmically rough at $d=2$.
For $d>2$, the fluctuations in the height are bounded and such a
surface is more or less flat, better called ``asymptotically flat''.
From the growth equation one can also derive the stationary
probability distribution for the height $h({\bf x})$ which takes the
form of a Boltzmann factor $P(h({\bf x})) \propto \exp[ -(\nu/D)
\int(\nabla h)^2 d^dx]$ resembling an equilibrium system at a
temperature given by $D=k_B T$.  This is the Einstein relation noise
should satisfy to recover equilibrium probability distribution.
Conversely, given a hamiltonian of the form $ \int (\nabla h)^2 d^dx$,
the equilibrium dynamics will be given by Eq. \ref{ew} with $D$
determined by the temperature.  Nevertheless, if we do not ascribe any
thermal meaning to $D$, Eq.  \ref{ew} is good enough to describe a
nonequilibrium dynamic process as well.  Such a nonequilibrium growth
will have many similarities with equilibrium processes, differing only
in the origin of the noise, e.g. the expected symmetry $h \rightarrow
-h$ with $\langle h\rangle =0$ in equilibrium will be preserved in the
nonequilibrium case also.  The growing surface with a correlation
$C({\bf x},t)=\mid x\mid^{(2-d)/2}\hat C (t/|x|^2)$ will be similar in
both cases for $d<2$.

A genuine nonequilibrium process will involve breaking the up-down
symmetry which in equilibrium follws from detailed balance.  It should
therefore be represented by a term involving even powers of $h$.  We
already saw that a constant current (zeroth power) does not add
anything new.  Since the origin in space or time or the position of
the basal plane should not matter, the first possible term is $(\nabla
h)^2$.  By looking at the geometry of a rough surface, it is easy to
see that such a term implies a lateral growth that would happen if a
deposited particle sticks to the first particle it touches on the
surface.  One gets the Kardar-Parisi-Zhang (KPZ) equation\cite{kpz86}
\begin{eqnarray}
\frac{\partial h}{\partial t}=\nu \nabla^2 h+\frac{\lambda}{2}(\nabla h)^2+
\eta({\bf x},t).
\label{eq:1}
\end{eqnarray}
As a consequence of its mapping to the noisy Burger's equation, to the
statistical mechanics of directed polymer in a random medium and other
equilibrium and nonequilibrium systems, the KPZ equation has become a
model of quite widespread interest in statistical mechanics.  Though
we focus on growth problems in this paper, the KPZ equation is also
applicable in erosion processes.

Taking a cue from the development in understanding the growth
phenomena through the KPZ equation, a vast class of simulational and
analytical models have evolved to explain different experimentally
observed growth processes.  Diverse technical tools ranging from
simulations with various dynamical rules to different versions of
renormalization group techniques, mode coupling theory, transfer
matrix techniques, scaling arguments have been employed to understand
kinetic roughening.  In this review we attempt to provide an overview
of this phenomenon of roughening of a growing surface. It is almost
beyond the scope of this review, to describe in detail various models
and their experimental relevance. Rather we focus our attention on a
few examples which may broadly represent a few different routes along
which research has continued.

The plan of this article is as follows. In the next section we focus
on the KPZ equation and its renormalization group description.  We
also point out the connection of the KPZ equation to some other
problems of physics.  In section III a more generalized growth
mechanism involving nonlocal interactions is presented.  Section IV is
devoted to the progress in understanding the roughening and super
roughening transitions which appear in a very distinct class of models
involving lattice pinning potential.

\section{KPZ equation and more}

Let us first look at the origin of the various terms in Eq. \ref{ew}
and \ref{eq:1}.  In both the equations, the noise term represents
random deposition, the fluctuation around the steady value. As already
mentioned a steady current can be removed from the equation by going
over to a moving frame.  The term involving second derivative of $h$
can represent either of two processes. It could be a surface tension
controlled diffusion process, in which a particle comes to the surface
and then does a random walk on the surface to settle at the minimum
height position thereby smoothening the surface.  An alternative
interpretation would be that there is desorption from the surface and
the process is proportional to the chemical potential gradient.  The
chemical potential of the particles on the surface cannot depend on
$h$ or gradient of $h$ because of its independence of the arbitrary
base or its tilt.  The chemical potential then is related to the
second derivative of $h$.  This also has a geometric meaning that
$\nabla^2 h$ is related to the local curvature.  The larger the
curvature, the higher is the chance to desorb because of a lesser
number of neighbours.  In the KPZ equation, the nonlinear term
represents lateral growth. The diffusion-like term can then be thought
of either (a) as an alternative that a particle coming to the surface
instead of sticking to the first particle it touches, deposits on the
surface and then diffuses, or (b) as a random deposition process with
desorption.  In either case, the noise term tends to roughen the
surface, the diffusion term, of whatever origin, smoothens it while
the nonlinear term leads to a laterally growing surface.  Even if the
smoothening linear term is not present, renormalization group or the
scaling argument indicates that such a term is generated on a large
length scale.

The KPZ equation has a special symmetry not present in the
Edwards-Wilkinson case.  This is the tilt symmetry (often called
Galilean invariance - a misnomer, though, in this context).  If we
tilt the surface by a small angle, then with a reparametrization
$h'=h+{\bf \epsilon} \cdot {\bf x}$ and ${\bf x}={\bf x}+\lambda
\epsilon t'$ and $t=t'$, the equation remains invariant for small
$\epsilon$. This transformation depends only on $\lambda$ the
coefficient of the nonlinear term and fails for $\lambda=0$.  Since
this tilt symmetry is to be maintained no matter at what lengthscale
we look at, $\lambda$ must be a renormalization group invariant.

Let us now perform a length rescaling analysis.  Under a change of
scale as $x\rightarrow bx$, $t\rightarrow b^z t$ and $h\rightarrow
b^{\chi} h$, KPZ equation transforms as
\begin{eqnarray}
b^{\chi-z}\frac{\partial h}{\partial t}=\nu b^{\chi-2}\nabla^2 h+
\frac{\lambda}{2} b^{2\chi-2} (\nabla h)^2+b^{-d/2-z/2}\eta,
\label{eq:2}
\end{eqnarray}
where the noise correlation has been used to obtain the scaling of the
noise term. Therefore under this scale transformation different
parameters scale as $\nu \rightarrow b^{z-2}\nu$, $D\rightarrow
b^{z-d-2\chi}D$ and $\lambda\rightarrow b^{\chi+z-2}\lambda$. For
$\lambda=0$, the equation remains invariant provided $z=2$ and
$\chi=(2-d)/2$. These are just the exponents one expects from the EW
model.  (Such surfaces with anisotropic scaling in different
directions like $x$ and $h$ are called self-affine.  ) Though we
cannot predict the exponents from Eq. \ref{eq:2} when $\lambda \neq
0$, it does tell us that a small nonlinearity added to the EW equation
scales with a scaling dimension $\chi+z-2$.  This term is always
relevant in one dimension, because it scales like $b^{1/2}$.  This
type of scaling argument also shows that no other integral powers of
derivatives of $h$ need be considered in Eq. \ref{eq:1} as they are
all irrelevant, except $(\partial h/\partial x)^3$ at $d=1$, which,
however, detailed analysis shows to be marginally irrelevant. Based on
this analysis, we reach an important conclusion that the
nonequilibrium behaviour in one dimension, and in fact for any
dimensions below two, would be distinctly different from the
equilibrium behaviour.  For dimensions greater than two, Edwards
Wilkinson or equilibrium surfaces, as already mentioned, are
asymptotically flat with $\chi=0, z=2$, and so, a small nonlinearity
is irrelevant because it will decay with $b$. In other words, the
growth in higher dimensions for small $\lambda$ would be very similar
to equilibrium problems because the EW model is stable with respect to
a small perturbation with nonlinearity.  The simple scaling argument
does not tell us if the nature of the surface changes for large
$\lambda$ for $d>2$, but an RG analysis shows that it does change.
That the nonequilibrium growth is always different in low dimensions
and in higher dimensions (greater than two), and that there will be a
dynamic phase transition from an equilibrium-like to a genuine
nonequilibrium behaviour, explains the source of excitement in this
minimal KPZ equation, in the last two decades.

If the nonlinear parameter $\lambda$ is to remain an invariant, i.e.
independent of $b$ in Eq. (\ref{eq:2}), then $\chi + z=2$, a relation
which need not be satisfied by the equilibrium growth.  It is this
relation connecting the two exponents of the scaling function of Eq.
\ref{scaling} that distinguishes nonequilibrium growth from
equilibrium, the former requiring one less exponent than the latter
one. We wonder if such an exponent relation is generally true for all
nonequilibrium systems.

Though we are far away from a complete understanding of all the
nuances and details of the KPZ equation, the renormalization group
analysis has been very successful in identifying different phases,
nature of phase transitions, and, in certain cases, relevant
exponents.  In brief, the various results obtained from
renormalization group analysis are as follows.  In one dimension and
for $d<2$, even a small nonlinearity, as already mentioned, being
relevant in the RG sense, leads to new values of roughening and
dynamic exponents, and is characterized by a RG fixed point.  Beyond
$d=2$, there is a phase transition demarcating two different types of
surfaces. A small nonlinearity is irrelevant around EW model and the
surface is almost flat with $\chi=0$ and $z=2$. A strong nonlinear
growth, however, drives the system to a different phase with rougher
surface where $\chi\neq 0$.  Several aspects of this phase transition
can be studied from RG but the strong $\lambda$ regime is still out of
reach, because of the absence of any RG fixed point.

The KPZ equation in $d=1$ has distinct nonequilibrium behaviour, and
the scaling behaviour is the same no matter how small or large
$\lambda$ is. More peculiar is the existence of a stationary
probability distribution of the height in one dimension which is the
same as for the linear EW model.  This is not just an accident but a
consequence of certain subtle relations valid only in one dimension.
We do not go into those issues here. The same stationary distribution
implies that the nonlinearity does not affect the stationary state
solution, and $\chi=1/2$.  The two models however differ in the
dynamic exponent which, in the case of KPZ growth, has to satisfy
$\chi + z =2$.  This leads to exact answer $z=3/2$.  Its significance
can be grasped if we compare various known cases.  For ballistic
motion, distance goes linearly with time so that the dynamic exponent
is $z=1$ while for diffusive motion or in quantum mechanics (e.g. a
nonrelativistic free quantum particle), $z=2$ as also the case for EW.
Here is an example where the nonequilibrium nature of the problem
leads to a completely new exponent connecting the scaling of space and
that of time.

\subsection{Dynamic renormalization group analysis}
A dynamic renormalization group analysis is a more general approach
applicable for dynamics which e.g. may be governed by the Langevin
equation for the appropriate dynamical variable. For our problem it is
easier to work in Fourier coordinates ${\bf q}$, and $\omega$
conjugate to space and time.  Long distance, long time implies ${\bf
  q} \rightarrow 0$ and $\omega\rightarrow 0$, and $q$ can be taken as
the inverse wavelength at which the height variable is probed. The
magnitude of wave vector ${\bf q}$ varies from $0$ to $\Lambda$ where
the upper cutoff is determined by the underlying microscopic length
scale like lattice spacing or size of particles etc.  In the Fourier
space, different Fourier modes in the linear EW model gets decoupled
so that each $h({\bf q},\omega)$ for each $({\bf q},\omega)$ behaves
independently.  It is this decoupling that allows the simple rescaling
analysis of Eq. \ref{eq:2} or dimensional analysis to give the correct
exponents.  For the KPZ equation the nonlinear term couples heights of
various wavelengths and therefore any attempt to integrate out the
large $({\bf q}, \omega)$ modes will affect $h$ with low values of
$({\bf q}, \omega)$.  This mixing is taken into account in the RG
analysis which is implemented in a perturbative way.  One thinks of
the noise and the nonlinear term as disturbances affecting the EW-like
surface.  If we know the response of such a surface to a localized
disturbance we may recover the full response by summing over the
disturbances at all the points and times.  However this disturbance
from the nonlinear term itself depends on the height, requiring an
iterative approach that generates successively a series of terms.  By
averaging over the noise, one then can compute any physical quantity.
At this stage only degrees of freedom with ${\bf q}$ in a small shell
$e^{-l} \Lambda<q<\Lambda$ is integrated out. In real space this
corresponds to integrating out the small scale fluctuation. The
contribution from this integration over the shell is absorbed by
redefining the various parameters $\nu$, $\lambda$ and $D$.  These are
the coupling constants for a similar equation as (\ref{eq:1}) but with
a smaller cutoff $\Lambda e^{-l}$.  A subsequent rescaling then
restores the original cutoff to $\Lambda$.  Following this procedure,
the flow equations for different parameters $\nu$, $D$, and $\lambda$
can be obtained\cite{medina}. Using the exponent identity predicted
from the Galilean invariance and the renormalization group invariance
of $\nu$, the flow equations for all the parameters can be combined
into a single flow equation for $\overline\lambda^2=\lambda^2D/\nu^3$
(with $\Lambda=1$).  This is the only dimensionless parameter that can
be constructed from $\lambda$, $\nu$, $D$, and $\Lambda$, and it is
always easier to work with dimensionless quantities.  Its recursion
relation is
\begin{eqnarray}
\frac{d\overline \lambda}{dl}=\frac{2-d}{2} \overline \lambda+
K_d\frac{2d-3}{4d}\overline\lambda^3,\label{flow1}
\end{eqnarray}
where $K_d$ is the surface area of a $d$-dimensional sphere divided by
$(2\pi)^d$.  The invariance of $\nu$ under RG transformation implies
$z=2-K_d\overline\lambda^2\frac{2-d}{4d},$ and the Galilean invariance
provides the value of $\chi=2-z$ once the value of $z$ is known. To be
noted here is that the dynamic exponent is different from 2 by a term
that depends on $\lambda$ coming from the renormalization effects.

A few very important features are apparent from (\ref{flow1}). From
the fixed point requirement $d\overline\lambda/dl=0$, we find that at
$d=1$, there is a stable fixed point $\overline \lambda^2=2/K_1$. At
this fixed point $z=3/2$ and $\chi=1/2$ supporting the results
predicted from the symmetry analysis.  At $d=2$, the coupling is
marginally relevant indicating a strong coupling phase not accessible
in a perturbation scheme. At $d>2$, the flow equation indicates two
distinct regimes, namely a weak coupling regime where
$\overline\lambda$ asymptotically vanishes leading to a flat EW phase
with $\chi=0,\ z=2$, and a strong coupling rough phase, the fixed
point of which cannot be reached by perturbation analysis.

Owing to this limitation of the renormalization group analysis based
on the perturbation expansion, the scaling exponents in this strong
coupling phase cannot be determined by this RG scheme.  Different
numerical methods yield $z=1.6$ at $d=2$. The phase transition
governed by the unstable fixed point of $\overline\lambda$ is well
under control with $z=2$ for all $d>2$.  To explore the strong
coupling phase, recently techniques like self-consistent mode coupling
approach, functional renormalization group etc have been employed, but
even a basic question whether there is an upper critical dimension at
which $z$ will again become 2 remains controversial.

\subsection{Relation with other systems}

The relation of the KPZ equation with other quite unrelated topics in
equilibrium and nonequilibrium statistical mechanics is impressive.
Here we provide a very brief account of these systems.

{\it Noisy Burgers equation:}
By defining a new variable 
${\bf v}=\nabla h$, we obtain an equation 
\begin{eqnarray}
\frac{\partial {\bf v}}{\partial t}=D\nabla^2 {\bf v}+\lambda {\bf v}\cdot 
\nabla v+{\bf f}({\bf x},t),
\end{eqnarray} 
where the noise term $f=\nabla \eta$.  The above equation represents
the 
noisy Burgers equation for vortex free ( $\nabla \times {\bf v}=0$)
fluid flow with a random force.  This equation is very
important in studies of turbulence.  The tilt invariance of the KPZ
equation turns out to be the conventional Galilean invariance for the
Burgers equation (for $\lambda=1$), and that's how the name stayed on.

{\it Directed polymer in a random medium:} A directed polymer, very
frequently encountered in different problems in statistical mechanics,
is a string like object which has a preferred longitudinal direction
along which it is oriented, with fluctuations in the transverse
direction.  The flux lines in type II or high T$_c$ superconductors
are examples of such directed polymers in 3 dimensions, while the
steps on a vicinal or miscut crystal surface or the domain walls in a
uniaxial two dimensional system are examples in two dimensions.  The
formal mathematical mapping to such objects follows from a simple
(Cole-Hopf) transformation of the KPZ equation using $ W({\bf
  x},t)=\exp[\frac{\lambda}{2\nu} h].  $ The Cole-Hopf transformation
linearizes the nonlinear KPZ equation and the resulting linear
diffusion equation (or imaginary time Schroedinger equation) is
identical to that satisfied by the partition function of a directed
polymer in a random potential.  For such random problems, one is
generally interested in the averages of thermodynamic quantities like
the free energy and we see that the noise averaged height $\langle
h({\bf x},t)\rangle$ gives the average free energy of a directed
polymer of length $t$ with one end at origin and the other end at
${\bf x}$.  This is a unique example of a system where the effect of
such quenched averaging of free energy can be studied without invoking
any tricks (like the replica method).  This has led to many important
results and enriched our understanding of equilibrium statistical
mechanics.  Recently, this formulation has been extended to study
details of the properties of the random system near the phase
transition point and overlaps in lower dimensions\cite{mezard,smpre}.
It turns out that one needs an infinite number of exponents to
describe the statistical behaviour of the configurations of the
polymer in the random medium\cite{sm96}.  We do not go into this issue
as this is beyond the scope of this article.

An interesting connection between the $1+1$ dimensional KPZ equation
and the equilibrium statistical mechanics of a two dimensional smectic
-A liquid crystals has been recently established by Golubovich and
Wang\cite{golub}.  This relationship further provides exact approach
to study the anomalous elasticity of smectic-A liquid crystals.

Apart from these, there are a number of other relations between KPZ
equation and kinetics of annihilation processes with driven diffusion,
the sine-Gordon chain, the driven diffusion equation and so on.

\subsection{Beyond KPZ }
\subsubsection{Conservation condition}
The situation encountered in Molecular-beam epitaxy (MBE) for growth
of thin films is quite different than the mechanism prescribed by the
KPZ equation\cite{barbasi}.  In MBE the surface diffusion takes place
according to the chemical potential gradient on the surface,
respecting the conservation of particles.  If the particle
concentration does not vary during growth, then a mass conservation
leads to a volume conservation and the film thickness is governed by
an underlying continuity equation
\begin{eqnarray}
\frac{\partial h}{\partial t}+\nabla \cdot {\bf j}=\eta,
\end{eqnarray}
where $j$ is the surface diffusion current which states that the
change of height at one point is due to flow into or out from that
point.  The current is then determined by the gradient of the chemical
potential, and since the chemical potential has already been argued to
be proportional to the curvature $\nabla^2 h$, the growth equation
thus becomes a simple linear equation involving $\nabla^4 h$ which,
like the EW model, is exactly solvable.  Taking into account the
effect of nonlinearity the full equation can be written as
\begin{eqnarray}
\frac{\partial h}{\partial t}=-\nabla^2[\nu \nabla^2 h+\frac{\lambda}{2}
(\nabla h)^2]+\eta(x,t),
\end{eqnarray}
where the noise correlation is $\langle \eta({\bf x},t)\eta({\bf
  x}',t')\rangle=2D\nabla^2 \delta({\bf x}-{\bf x}')\delta(t-t')$, if
the noise also maintains conservation (if it originates from the
stochasticity of diffusion) or would be the same white noise as in the
KPZ equation if the noise is from random deposition.  It goes without
saying that the exponents are different from the EW model even for the
linear theory. The invariance of $\lambda$ in this case leads to a
different relation between $\chi$ and $z$. At the dimension of
physical interest $d=2$, this 
growth equation leads to an enhanced roughness than the KPZ case and
may explain the results of experiments of high temperature MBE.

\subsubsection{Quenched noise}

A different type of generalization of the KPZ equation was to explore
the motion of domain walls or interfaces in a random medium.  In this
case, the noise is not explicitly dependent on time but on the spatial
position and the height variable.  Such a noise has been called
quenched noise because the noise is predetermined and the interface or
the surface moves in this random system.  The simple features of the
KPZ equation and the EW model are lost.  Functional renormalization
group analysis and numerical studies attempted to clarify the question
of universality classes and details of dynamics in such cases.  The
important concept that emerged in this context is the depinning
transition so that the surface remains pinned by the randomness until
the drive exceeds a certain critical value. Interface depinning is an
example of a nonequilibrium phase transition.  The velocity of the
surface near this depinning transition also has critical like
behaviour with long range correlations. Below the threshold, the
dynamics is sluggish, while just above the  threshold , the
velocity is in general not proportional to the drive but obeys a power
law with a universal exponent.  For a very strong drive (or large
velocity of the interface) the moving surface encounters each site
only once, and therefore the noise is effectively like a space-time
dependent noise rather than the quenched one.  The nature of the
surface would then be like KPZ.

\subsubsection{Coloured noise}
In the previous section we discussed the KPZ equation with white
noise.  If the noise is coloured in the sense that there is
correlation in space or time or both, the universal behaviour, the
phase transitions and the properties are different but still can be
studied by the same RG technique. Several aspects of the problem
especially the role of noise correlation have been
explored\cite{hayat}.

All of the above seem to suggest that if there is no conservation law,
then the KPZ equation is the equation to describe any nonlinear or
nonequilibrium growth process and all phenomena can be put in one of
the known universality classes.  However, experimentally KPZ exponents
seem to elude us so far\cite{colloid,barbasi,tim}.  Since results are
known exactly in one dimension, special one dimensional experiments
were conducted like paper burning, interface motion in paper, colloid
suspension etc, but KPZ exponents have not been seen.  In the colloid
experiment\cite{colloid} the surface formed by the depositing colloids
on the contact line $(d=1)$ between the colloid latex film and a glass
slide was measured from video images.  This method yields $\chi=.71$
but cannot determine the dynamic exponent.  A recent
analysis\cite{cloud} of tropical cumulus cloud in the upper
atmosphere, from satellite and space shuttle data from 0.1 to 1000 Km,
seems to agree with the KPZ results in $d=2$.

\section{Kinetic roughening with nonlocality}

In spite of a tremendous conceptual and quantitative success of KPZ
equation in describing the nonequilibrium growth mechanism, the
agreement with experimentally observed exponents is rather
unsatisfactory. One wonders whether there is any relevant perturbation
that drives the systems away from the KPZ strong coupling
perturbation.  One goal of this section is to point out that indeed
there can be longrange interactions that may give rise to non-KPZ
fixed points.

Many recently studied systems involving proteins, colloids, or latex
particles the medium induced interactions are found to play an
important role\cite{nonlocal1}. This nonlocal interaction can be
introduced by making a modification of the nonlinear term in the KPZ
equation. Taking the gradient term as the measure of the local density
of deposited particles, the long range effect is incorporated by
coupling these gradients at two different points.  The resulting
growth equation is a KPZ-like equation with the nonlinear term
modified as \cite{mbprl} $\frac{1}{2}\int d{\bf r}' {\cal V}(r') $
$\nabla h ({\bf r}+{\bf r}',t)\cdot \nabla h({\bf r}-{\bf\ r}',t)$.
For generality we take ${\cal V}(r')$ to have both short and long
range parts with a specific form in Fourier space as ${\cal
  V}(k)=\lambda_0+\lambda_\rho k^{-\rho}$ such that in the limit
$\lambda_\rho\rightarrow 0$ KPZ results are retrieved. The aim is to
observe whether the macroscopic properties are governed by only
$\lambda_0$ and hence KPZ like or the behavior is completely different
from KPZ due to the relevance of $\lambda_\rho$ around the KPZ fixed
points.

A scaling analysis as done in Eq. (\ref{eq:2}) clearly indicates
different scaling regimes and the relevance of $\lambda_0$ and
$\lambda_\rho$ for $d<2$ at the EW fixed point.  For any
$\lambda_\rho$($\neq 0$) with $\rho>0$, the local KPZ theory (i.e.
$\lambda_\rho=0$ and $\chi+z=2$) is unstable under renormalization and
a non-KPZ behavior is expected. For $2<d<2+2\rho$, only
$\lambda_{\rho}$ is relevant at the EW fixed point. The exponents of
the non-KPZ phases can be obtained by performing a dynamic
renormalization group calculation\cite{mbprl}.  By identifying the
phases with the stable fixed points, we then see the emergence of a
new fixed point where the long range features dominate
$(\chi+z=2+\rho)$.  Most importantly, at $d=2$, the marginal relevance
of $\overline \lambda$ 
is lost and there is a stable fixed point (LR) for $\rho>.0194$.
 
On the experimental side, there are experiments on colloids with
$\chi=0.71$ which is the value also obtained from paper burning
exponents. For colloids, hydrodynamic interactions are important.
Similar longrange interactions could also play a role in paper burning
experiment due to the microstructure of the paper. With this $\chi$
our exponents suggest $\rho=-0.12$ at $d=1$ at the long range fixed
point. Further experiments on deposition of latex particles or
proteins yielding the roughness of growing surface have not been
performed. Probably such experiments may reveal more insights on this
growth mechanism.

More recently, the effect of coloured noise in presence of nonlocality
has been studied\cite{achat} and the nature of the phases and the
various phase transitions clarified.  A conserved version of the
nonlocal equation has also been considered and it shows rich
behaviour\cite{kim}.

\section{Roughening transition in nonequilibrium}

It is interesting to study the impact of equilibrium phase transitions
on the nonequilibrium growth of a surface. This is the situation
observed experimentally in growth of solid ${}^4He$ in contact with
the superfluid phase\cite{nozier}. There is an equilibrium roughening
transition at $T_R=1.28 K$. For $T>T_R$ the growth velocity is linear
in the driving force $F$ (chemical potential difference), but for
$T<T_R$ the velocity is exponentially small in the inverse of the
driving force. For infinitesimal drive , the mobility which is the
ratio of the growth velocity and $F$ vanishes with a jump from a
finite value at the transition.  With a finite force the transition is
blurred and the flat phase below $T_R$ in equilibrium becomes rougher
over large length scale.

The equilibrium roughening transition is an effect of discrete
translational symmetry of the lattice. The equilibrium dynamics in
this case is essentially governed by the Langevin equation
\begin{eqnarray}
\frac{\partial h}{\partial t}=K \nabla^2 h({\bf r},t)
 -V\sin[\frac{2\pi}{a} \ h(r,t)]+\zeta(r,t),
 \label{equilang}
\end{eqnarray}
where the sin term favours a periodic structure of spaoing $a$.
Extensive investigations have been done on this equilibrium model.  At
low temperature this periodic potential is relevant and it ensures
that minimum energy configuration is achieved when $\phi$ is an
integer multiple of lattice periodicity. In this phase the surface is
smooth and the roughness is independent of length.  In the high
temperature phase the equilibrium surface is thermally rough and the
roughness is logarithmic
\begin{eqnarray}
 C(L,\tau)\sim \ln[Lf(\tau/L^z)].
\end{eqnarray}  
The critical point is rather complicated and goes by the name of
Kosterlitz-Thouless transition, first discussed in the context of
defect mediated transitions in two dimensional XY magnets\cite{chui}.
 
For a nonequilibrium crystal growth problem, one needs to introduce
the KPZ nonlinear term in (\ref{equilang}).  There is no longer any
roughening transition.  The fact that away from equilibrium the
roughening transition is blurred is manifested by the domination of
the nonlinear term and the suppression of the pinning potential in the
asymptotic regime\cite{rost}.
 
A very nontrivial situation arises when the surface contains quenched
disorder which shifts the position of the minima of the pinning
potential in an arbitrary random fashion\cite{tsai,scheidl}. In this
case there is a new phase transition which is drastically different from
the equilibrium roughening transition.  This transition is called
super roughening.  Above the transition temperature i.e. for
$T>T_{sr}$, the surface is logarithmically rough as it is in the high
temperature phase of the pure problem. However in the low temperature
phase i.e.  for $T<T_{sr}$, the surface is no longer flat and is even
rougher than the high temperature phase.  Recent numerical treatments
suggest that the surface roughness behaves as $(\ln L)^2$.  In the
nonequilibrium situation the linear response mobility vanishes
continuously at the transition temperature unlike the jump
discontinuity in the pure case.  A general treatment with a correlated
disorder elucidates the connection between the roughening and super
roughening transition and one observes that the roughening turns into
a super roughening transition if the disorder correlation decays
sufficiently fast. Away from equilibrium, the super roughening
transition is essentially dominated by the KPZ nonlinearity and
instead of the logarithmic roughness, an asymptotic power law behavior
of the roughness is found over all temperature range.
 
In a similar situation in the nonequilibrium case, one needs to study
the role of the KPZ nonlinearity with long range disorder
correlation\cite{sm97}.  A functional renormalization scheme with an
arbitrary form of the disorder correlation turns out to be useful,
though a detailed solution is not available.  It is found that the
flow of the KPZ nonlinearity under renormalization, with power law
form of the disorder correlation, is such that it decays with length.
This implies that nonequilibrium feature does not set in over a
certain length scale. Over this scale one would then expect usual
roughening transition. However there is a generation of a driving force
due to the nonlinearity, and the growth of this force with length
scale would invalidate use of perturbative analysis.  For large length
scales, one expects a KPZ type power law roughness of the surface.
Nevertheless, the initial decay of the nonlinearity with the length
scale due to the long range correlation of the disorder is an
interesting conclusion that seems to be experimentally detectable.

\section{Remarks}
In this brief overview, we attempted to focus on the difference
between equilibrium and nonequilibrium growth problems with an
emphasis on the scaling behaviour and renormalization group approach.
Many details with  references to pre-1995 papers can be found in Ref.
\cite{tim,barbasi}, which should be consulted for more detailed
analysis.  Though the success story of the KPZ equation is rather
impressive, there are still many unresolved, controversial issues.  In
fact for higher dimensions, the behaviour is not known with as much
confidence as for lower dimensions. Developments in this direction are
awaited.

{\bf Note added in Proof}

(1) The growth mechanism of metal-organic films deposited by the
Langmuir-Blodgett technique has been studied in Ref. 23 by X-ray
scattering and atomic force microscopy.  The results have been
interpreted by a combination of 1-dimensional EW equation (Eq. 4 ) and
2-dimensional linear conserved equation (Eq. 10) with conserved noise.

(2) For effects of nonlocality in equilibrium critical dynamics, see
Ref. 24.

\end{multicols}
\end{document}